\documentclass[twocolumn,english,aps,prb,showpacs,superscriptaddress]{revtex4}
\usepackage[T1]{fontenc}
\usepackage[latin9]{inputenc}
\usepackage{verbatim}
\usepackage{graphicx}
\usepackage{multirow}
\usepackage{babel}

\begin{document}

\title{A First-Principles Study of Defects and Adatoms in Silicon Carbide Honeycomb Structures}

\author{E. Bekaroglu}
\affiliation{UNAM-Institute of Materials Science and
Nanotechnology, Bilkent University, Ankara 06800, Turkey}
\author{M. Topsakal}
\affiliation{UNAM-Institute of Materials Science and
Nanotechnology, Bilkent University, Ankara 06800, Turkey}
\author{S. Cahangirov}
\affiliation{UNAM-Institute of Materials Science and
Nanotechnology, Bilkent University, Ankara 06800, Turkey}
\author{S. Ciraci}
\email{ciraci@fen.bilkent.edu.tr} \affiliation{UNAM-Institute of
Materials Science and Nanotechnology, Bilkent University, Ankara
06800, Turkey} \affiliation{Department of Physics, Bilkent
University Ankara 06800, Turkey}
\date{\today}

\begin{abstract}
We present a study of mechanical, electronic and magnetic
properties of two dimensional (2D), monolayer of silicon carbide
(SiC) in honeycomb structure and its quasi 1D armchair nanoribbons
using first-principles plane wave method. In order to reveal
dimensionality effects, a brief study of 3D bulk and 1D atomic
chain of SiC are also included. Calculated bond-lengths, cohesive
energies, charge transfers and band gaps display a clear
dimensionality effect. The stability analysis based on the
calculation of phonon frequencies indicates that 2D SiC
monolayer is stable in planar geometry. We found that 2D SiC
monolayer in honeycomb structure and its bare and hydrogen
passivated nanoribbons are ionic, non-magnetic, wide band gap
semiconductors. The band gap is further increased upon self-energy
corrections. The mechanical properties are investigated using the
strain energy calculations. The effect of various vacancy defects, adatoms and
substitutional impurities on electronic and magnetic properties in
2D SiC monolayer and in its armchair nanoribbons are also
investigated. Some of these vacancy defects and impurities, which
are found to influence physical properties and attain magnetic
moments, can be used to functionalize SiC honeycomb structures.
\end{abstract}

\pacs{73.22.-f, 75.75.+a, 63.22.-m}
\maketitle

\section{introduction}

Owing to its exceptional thermal and physical properties,\cite{sicbook} silicon carbide (SiC) is a material, which is
convenient for high temperature and high power device
applications. Because of its wide band gap, SiC bulk structure has
been a subject of active study in optical and optoelectronic
research. Unlike the polymorphs of carbon, SiC is a polar
material. In spite of the fact that both constituents of SiC are
Group IV elements, charge is transferred from Si to C  due to
higher electronegativity of C relative to Si atom.

Bulk SiC has six commonly used  stacking configurations denoted as
3C (zincblende), 2H (wurtzite), 4H, 6H, 15R and 21R. Lubinsky \textit{et al.}\cite{lubinsky} reported optical data related with indirect
transitions, dielectric function and reflectivity of 3C SiC using
first-principles Hartree-Fock-Slater method. A more comprehensive
study\cite{ching} using OLCAO (orthogonalized linear combination
of atomic orbitals) method comprises the calculations of lattice
constants, electronic band structure and optical properties of all
six stacking configuration of SiC.

As for SiC in lower dimensionality, SiO$_2$ coated  SiC nanowires
\cite{nwexp} were synthesized and showed favorable photocatalytic
behavior. A theoretical work on hydrogen passivated SiC
nanowires\cite{nwth} provided the energy bands both using local
density approximation within Density Functional Theory (DFT) and
$sp^{3}s^{*}$ LCAO tight binding (TB) methods. SiC-ZnS core-shell
structures were also fabricated.\cite{coax} Zincblende SiC
nanoparticles were synthesized by carbothermal reduction
method.\cite{carbo} Band gap of zincblende nanoparticles were
estimated to be around 3 eV from photoluminesence measurements. With
a similar carbothermal method, microribbons\cite{micror} with
widths in the range of 500 nm - 5 $\mu$m and thickness of 50-500
nm were synthesized.

SiC is frequently used as a substrate to grow other
materials.\cite{asasub1,asasub2} Few layers of graphene was also
grown on SiC.\cite{graonsic} SiC clusters (Si$_n$C$_n$, n=1-10)
were investigated\cite{cluster} using DFT. With the aim of
developing a material for future nanoelectronic applications,
binding energy, HOMO-LUMO gap, Mulliken charge, vibrational
spectrum  and ionization potential of Si$_n$C$_n$ clusters are
revealed.

Earlier, planar honeycomb structure of graphite was exfoliated and
its physical properties were analyzed.\cite{novo,zhang,berger}
While graphene is a strictly planar crystal, the planar honeycomb structure of
Si is unstable, but it is stabilized through puckering.\cite{silicene} Since the honeycomb structure is common
to both C and Si, one expects that stable 2D SiC in honeycomb
structure can be synthesized.

In this paper, a comprehensive analysis of the atomic, electronic
and magnetic properties of 2D monolayer of SiC honeycomb structure
and its bare and hydrogen passivated armchair nanoribbons
(A-SiCNR) are carried out using first-principles calculations. In
spite of the fact that 2D SiC monolayer is not synthesized yet,
this study demonstrates its stability based on reliable
theoretical methods. Furthermore, various mechanical, electronic
and magnetic properties are revealed. We started with the
discussion of 3D zincblende and wurtzite crystals, as well as SiC
atomic chain as an ultimate 1D system; we presented an analysis of
optimized atomic structures with corresponding  phonon dispersion
curves and electronic energy band structures and effective
charges.  Then we provided an extensive analysis of 2D and
quasi-1D (nanoribbon) SiC in terms of the optimized atomic
structures and their stability, electronic and magnetic
structures. We revealed elastic constants, such as in-plane
stiffness and Poisson's ratio. Having obtained the results
for 1D, 2D and 3D structures, we presented a comprehensive
discussion of dimensionality effects. Then we investigated the
effect of vacancy defects (such as Si and C vacancy, Si+C vacancy
and C-Si antisite defect) on the electronic and magnetic
properties of single layer SiC and its armchair nanoribbons.
Furthermore, we showed that  SiC can be functionalized through
adsorption of a foreign atom to the surface of 2D SiC or through
substitution of either C or Si with a foreign atom. It is revealed
that 2D SiC and its ribbons provide unusual physical properties,
which are absent in 3D SiC crystals. For example, while various allotropic forms of SiC
including its honeycomb structures are normally non-magnetic
semiconductors, a Si vacancy gives rise to spin polarization.
Significant variation of the band gap of narrow A-SiCNR's with their widths may be
crucial in designing optoelectronic nanodevices.

\section{Model and Methodology}

We have performed first-principles plane wave calculations within
DFT using PAW potentials.\cite{paw} The exchange correlation
potential has been approximated by Generalized Gradient
Approximation (GGA) using PW91\cite{pw91} functional both for
spin-polarized and spin-unpolarized cases. For the sake of
comparison, the calculations are also carried out using different
potentials and exchange-correlation approximations. All structures
have been treated within the supercell geometry using the periodic
boundary conditions. A plane-wave basis set with kinetic energy
cutoff of 500 eV has been used. A vacuum spacing of 12 \AA{} hinders the interactions between SiC monolayers in adjacent supercells. In the self-consistent structure optimizations, the
Brillouin zone (BZ) is sampled by, respectively
(5$\times$5$\times$5), (11$\times$11$\times$1) and
(11$\times$1$\times$1) special \textbf{k}-points for 3D bulk, 2D
honeycomb and 1D nanoribbons of SiC. Further relaxation is made
with (11$\times$11$\times$11), (31$\times$31$\times$1) and
(25$\times$1$\times$1) special \textbf{k}-points in order to find
the final structure.  All atomic positions
and lattice constants are optimized by using the conjugate
gradient method, where the total energy and atomic forces are
minimized. The convergence for energy is chosen as 10$^{-5}$ eV
between two steps and the maximum Hellmann-Feynman forces acting
on each atom is less than 0.04 eV/\AA{} upon ionic relaxation. The
pseudopotentials corresponding to 4 valence electrons of Si
(Si:$3s^{2}$ $3p^{2}$) and C (C:$2s^{2}$ $2p^{2}$) are used.
Numerical plane wave calculations are performed by using
VASP.\cite{vasp1,vasp2} Part of the calculations have also been
repeated by using  SIESTA\cite{siesta} software. The cohesive
energy of any SiC structure is found as $E_C={E_{T}[SiC]} -
E_{T}[Si] - E_{T}[C]$ in terms of the optimized total energy of
any SiC structure, and the spin-polarized total energies of free Si and C atoms,
all calculated in the same supercell using the same parameters.
Phonon calculations were carried out using PHON program\cite{alfe}
implementing force constant method. \textit{GW$_{0}$} calculations\cite{gw}
are again handled by VASP.

\section{1D atomic chain and 3D bulk crystal of SiC}

In this section, we present a brief discussion of 1D SiC atomic
chain and 3D bulk crystal based on our structure optimized total
energy and phonon spectrum calculations. Studies on SiC bulk lattice and atomic
chains already exist in the literature.\cite{ching,lubinsky,enginch} However, our
purpose is to carry out calculations with same parameters as used
in 2D monolayer SiC honeycomb structure and provide a consistent
comparison of dimensionality effects.

\begin{figure}
\includegraphics[width=8cm]{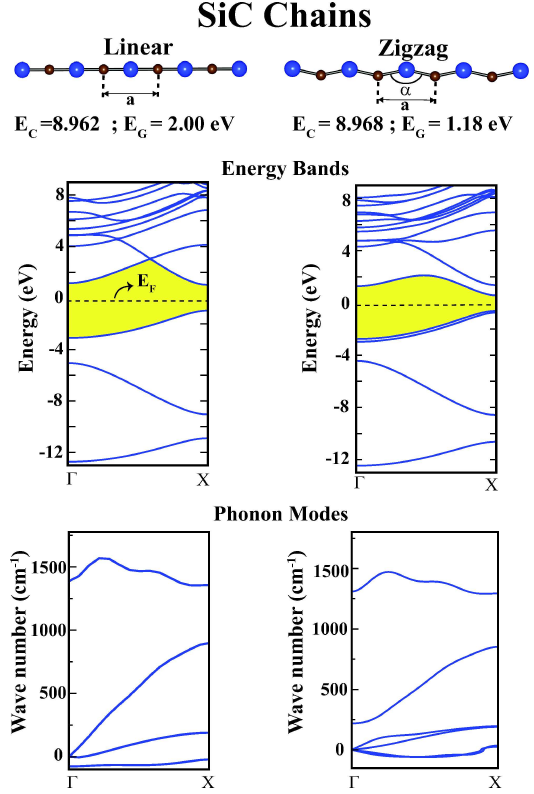}
\centering{}
\caption{(Color online) Atomic structures, electronic energy bands
and dispersion of phonon modes of linear and wide angle zigzag SiC
atomic chains. $E_{C}$ and $E_{G}$ are cohesive and band gap
energies, respectively. Si and C atoms are shown by blue/large and
brown/small balls, respectively. Zero of the energy is set to the Fermi energy, $E_{F}$. }

\label{fig:Chain}
\end{figure}

\subsection{1D SiC Chains}
Earlier, the first theoretical study of atomic chains of Group IV and III-V binary
compounds were reported by Senger \textit{et al.}\cite{enginch} They
examined SiC atomic chain as a function of lattice parameter and
found that the wide zigzag atomic chain of SiC with bond angle of
$\sim$147$^{\circ}$ is energetically more favorable than the
linear and narrow angle zigzag chains. Present  calculations find
that the atomic chains of SiC are non-magnetic. Calculated
structural parameters, cohesive energies, band gap and phonon
modes of linear and zigzag atomic chains, which are relevant for
the present study are given in Table ~\ref{table: chain}. The
charge transfer from Si to C is calculated to be $\delta$q= 2.28
electrons using the Bader analysis.\cite{bader} Phonon modes
calculated with force constant method have imaginary frequencies.
In Fig.~\ref{fig:Chain} two acoustical and two optical branches of
linear SiC chain have imaginary frequencies. Also wide angle
zigzag SiC chain has one optical and one acoustical branches with
imaginary frequencies. These results indicate that free standing
SiC chains are not stable. We note that carbon and BN atomic
chains are found to be stable and have linear
structure.\cite{enginch,tongay2,tongay3} Stability of linear chain structure is assured by $\pi$-bonding between adjacent atoms.

\begin{widetext}
\begin{table}
\caption{Si-C bond length, $d$; lattice constant, $a$; bond angle,
$\alpha$; charge transfer from Si to C, $\delta q$; band gap,
$E_{G}$; and cohesive energy, $E_{C}$ values for two different types of
SiC chains.} \label{table: chain}
\begin{center}
\begin{tabular}{ccccccc}
\hline  \hline Type  &  \textit{d} (\AA{})& \textit{a} (\AA{}) &
$\alpha$ (degree) & $\delta q$ (e) & $E_{G}$ (eV) & $E_{C}$ (eV)
\\ \hline \hline
Linear & 1.649 & 3.298 & 180 & 2.28 & 2.00 & 8.962         \\
\hline Wide Zigzag & 1.673 & 3.268  & 155.2 & 2.15 & 1.18 & 8.968
\\ \hline
\end{tabular}
\end{center}
\end{table}
\end{widetext}

\begin{figure}
\centering
\includegraphics[width=8cm]{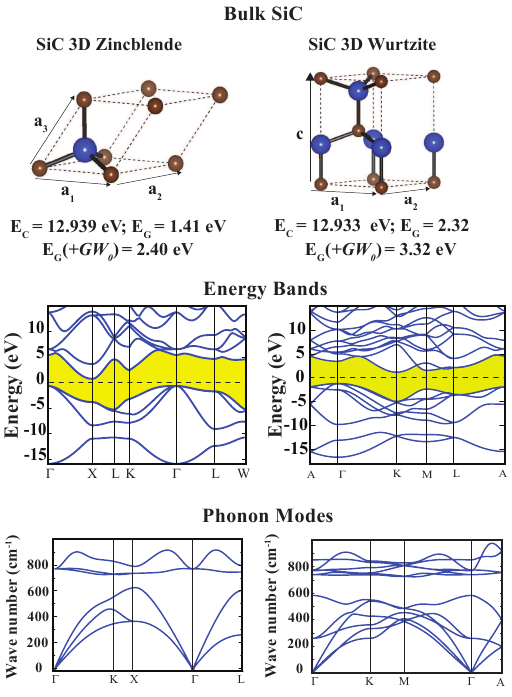}
\caption{(Color online) Optimized atomic structure with relevant
structural parameters, corresponding energy band structure and
frequencies of phonon modes of 3D bulk SiC in zincblende and
wurtzite structures. Zero of energy of the band structure is set
at the Fermi level, and band gap is shaded.}

\label{fig:Bulk}
\end{figure}

\subsection{3D SiC Crystals}

Our work on bulk SiC includes wurtzite (wz) and zincblende (zb)
structures. Atoms in wz- and zb-SiC are four fold coordinated
through tetrahedrally directed \textit{$sp^3$-}orbitals.
Calculated structural parameters, cohesive energies, energy band
structures and phonon modes are given in Fig.~\ref{fig:Bulk}.
Zincblende SiC structure in $T_d$ symmetry has cubic lattice
constants, $a_{1}=a_{2}=a_{3}= 3.096$ \AA{}. Si-C bond distance
\textit{d} is 1.896 \AA{}. Each Si (C) is connected to its four
nearest neighbor C (Si); four Si-C bonds are equal. Charge transfer
from Si to C is $\delta q$= 2.59 electrons calculated via Bader
analysis.\cite{bader}  While the GGA band gap is 1.41 eV, it
increases to 2.40 eV after \textit{GW$_{0}$} corrections. As for wz-SiC
crystal, the hexagonal lattice constants of the optimized
structure in equilibrium are $a_{1}=a_{2}= 3.091 $ \AA{}, $c/a =
1.642$. The small deviation of $c/a$ from the ideal value of 1.633
imposes a slight anisotropy on the lengths of tetrahedrally
directed Si-C bonds. While the length of three short bonds is
1.893 \AA, the fourth bond is slightly longer and has the length
of 1.907 \AA. Charge transfer from Si to C is $\delta q$= 2.63
electrons. The GGA band gap is  2.36 eV, but it increases to 3.32
eV after \textit{GW$_{0}$} correction. The calculated structural parameters
and energy band gaps are in reasonable agreement with the earlier
calculations and experimental measurements.\cite{bulkgap,ching} In
particular, the band gap values of 2.39 and 3.33 eV for zb and wz
SiC, respectively are in excellent agreement with the present
\textit{GW$_0$} corrected values.

The frequencies of phonon modes and their dispersions  are calculated for zb and wz
crystals by direct (or force constant)
method.\cite{alfe} At long wavelengths near the $\Gamma$ point, the electric field due to dipoles is critical for phonon modes. This effect lifts the degeneracy
between longitudinal and transverse optical modes. However the splitting (known as LO-TO splitting) cannot be observed with the direct method used in
the present study. Therefore, in Fig.~\ref{fig:Bulk} the highest
and second highest optical branches become degenerate at $\Gamma$
point. Present results are in agreement with earlier phonon
calculations.\cite{zbphonon,wzphonon}

\section{2D SiC honeycomb structure}
\begin{widetext} %
\begin{table}
\caption{Si-C bond length, \textit{d}; lattice constant, \textit{a}; band gap, $E_{G}$; band gap corrected by \textit{GW$_{0}$}, $E_{GW_{0}}$; cohesive energy, $E_{C}$ values
for 2D monolayer of SiC in honeycomb structure calculated with different
potentials}
\label{table: 2d}
\centering{}\begin{tabular}{ccccccc}
\hline
Potential  &  & \textit{d} (\AA{}) & \textit{a} (\AA{})  & $E_{G}$  & $E_{G}(+GW_{0})$ & $E_{c}$ (eV) \tabularnewline
\hline
\hline
PAW+GGA  &  & 1.786  & 3.094  & 2.530  & 3.90 & 11.944 \tabularnewline
\hline
PAW+LDA  &  & 1.770  & 3.070  & 2.510  & - & 13.542 \tabularnewline
\hline
US+GGA  &  & 1.776  & 3.079  & 2.542  & - & 11.973 \tabularnewline
\hline
US+LDA  &  & 1.759  & 3.048  & 2.532  & - & 13.471 \tabularnewline
\hline
\end{tabular}
\end{table}
\end{widetext}

\begin{figure*}
\includegraphics[width=17cm]{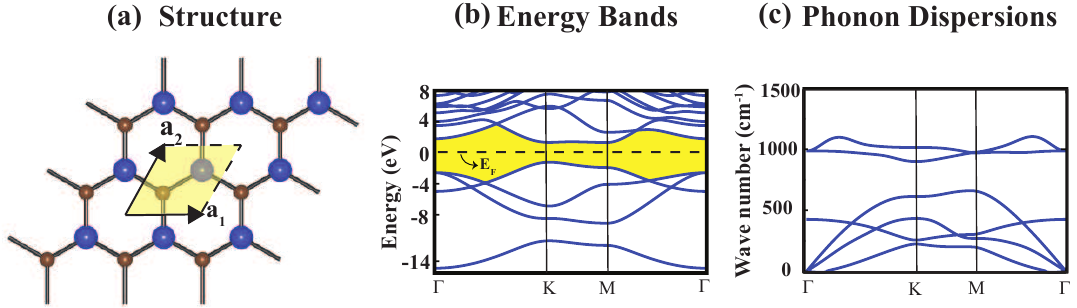}
\caption{(Color online) (a) atomic structure, (b) energy band
structure and (c) phonon modes of 2D SiC in honeycomb structure.
Large/blue and small/brown balls indicate Si and C atoms
respectively. The primitive unit cell is delineated. The zero of
energy in the band structure is set to the Fermi level.}
\label{fig: 2D}
\end{figure*}

Two-dimensional monolayer of SiC hexagonal
structure of SiC is optimized using periodically repeating
supercell having 12 \AA~ spacing between SiC planes. The minimum
of total energy occurred when Si and C atoms are placed in the
same plane forming a honeycomb structure. The magnitude of the
Bravais lattice vectors of the hexagonal lattice is found to be
$a_{1}=a_{2}$= $3.094$ \AA{} (see Fig.~\ref{fig: 2D}), and the Si-C
bond length to be $d=$1.786 \AA. The planar structure of 2D SiC is
tested by displacing Si and C atoms arbitrarily from their
equilibrium positions by 0.5 \AA{} and then reoptimizing the
structure. Upon optimization, the displaced atoms  returned to
their original positions in the same plane implying the stability
of planar structure. Further details on the stability of this
structure will be provided with phonon calculations at the end of
this section. Two dimensional monolayer SiC is found to be a
semiconductor with a band gap of 2.53 eV within GGA. Furthermore,
in Table ~\ref{table: 2d}, we present lattice constant, bond
length, cohesive energy, energy gap values of 2D SiC monolayer
calculated with different potentials. Since DFT usually
underestimates the band gap of semiconductors, we also corrected
the GGA band gap using \textit{GW$_{0}$} correction and found it to be 3.90
eV. The charge transfer from Si to C in 2D SiC is calculated to be
$\delta$q= 2.53 electrons. The Si-C bond length and the band gap values in the first row obtained by using GGA+PAW are in agreement with earlier DFT calculations.\cite{cinsic}

In addition to structural and electronic properties, we also
investigated the mechanical properties of 2D
SiC.\cite{hasan,prlprb,kopma} One can specify the mechanical
properties of SiC honeycomb sheet in terms of uniaxial strain,
$\epsilon = \Delta c/ c$, $c$ being the lattice constant;
Poisson's ratio $\nu=\epsilon_{trans}$/$\epsilon_{axial}$ and
in-plane stiffness,
$\textit{C}=\frac{1}{A_{0}}(\frac{\partial^{2}E_{s}}{\partial\epsilon^{2}})$.
Here $A_{0}$ is the equilibrium area of the system and $E_{s}$ is
the strain energy calculated by subtracting the total energy of
the strained system from the equilibrium total energy. To
calculate the elastic constants of monolayer SiC honeycomb
structure we switch to rectangular unit cell and consider a large
(8$\times$4) supercell comprising 32 primitive unit cells. In the
harmonic region $E_{s}(\epsilon)$ is first calculated on a 2D
grid. The numerical data is then fitted to the expression,
$E_{s}=A_{1}\epsilon_{x}^{2}+A_{2}\epsilon_{y}^{2}+A_{3}\epsilon_{x}\epsilon_{y}$;
where $\epsilon_{x}$ and $\epsilon_{y}$ are the small strains
along $x$- and $y$-directions. In the harmonic region, as a result of the isotropy
$A_{1}=A_{2}$. More details concerning the calculation of
\textit{C} and $\nu$ can be found in
references.\cite{prlprb,kopma} The calculated in-plane stiffness
of SiC honeycomb structure is found to be 166 $(J/m^{2})$. This is
almost half of the in-plane stiffness of graphene (namely 335
$J/m^{2}$), but more than twice the in-plane stiffness of silicene
(62 $J/m^{2}$). Also the Poisson's ratio of SiC is calculated to
be 0.29. Consequently, 2D SiC monolayer is a stiff material, but
less stiff than graphene and BN having similar honeycomb
structures.

Similar to 1D and 3D SiC, frequencies of phonon modes of and the
dispersions $\Omega$(\textbf{k}) of 2D monolayer of SiC in planar
geometry are calculated using direct method.\cite{alfe} Forces were found by displacing a single atom
in a 7$\times$7$\times$1 supercell. We use a small displacement in
order to stay in the harmonic region. We increased default grids
used by VASP until calculations converge. The lowest acoustical
mode, which is called as the out of plane, ZA mode is vulnerable to
instability. In rough meshes this mode gets imaginary frequencies
near the $\Gamma$-point, but it can be overcome by refining the
mesh along the z-axis (perpendicular to plane) as much as
possible. This way, force in that direction is calculated more
rigorously. Since the frequencies of all modes are positive in BZ, it
is concluded that planar, 2D SiC monolayer in honeycomb structure
is stable.

\subsection{Dimensionality effects}

In Table ~\ref{table: comparison}, we compare the variation of the
effective charge on Si and C atoms, namely $Z^{*}_{Si}$ and
$Z^{*}_{C}$ respectively; charge transfer from Si to C, $\delta q
= 4-Z^{*}_{Si}$; Si-C bond length $d$; lattice constant; energy
band gap, \textit{GW$_0$} corrected band gap and cohesive energy per Si-C,
calculated for SiC in different dimensionalities. It should be
noted that the length of Si-C bonds of 2D SiC honeycomb structure
is smaller than that in the 3D bulk (wz, zb) crystals, but larger
than that in zigzag atomic chains. Here we see that the
dimensionality effect is reflected to the strength of the bonding
through $sp^{n}$ hybridization, where $n$ coincides with the
dimensionality. While $sp^{2}$ hybrid orbitals of 2D planar
honeycomb structure form stronger bonds than tetrahedrally
coordinated $sp^{3}$ orbitals of 3D bulk, they are relatively
weaker than $sp$ hybrid orbitals of 1D chain. Accordingly, $d$ is
shortest in 1D chain, and longest in 3D zb structures, and is
intermediate in 2D monolayer. The cohesive energy $E_{C}$
increases with dimensionality, since the number of nearest
neighbors increases. Effective charge or charge transfer between
cation and anion also varies with dimensionality. For example, the
charge transfer calculated with Bader analysis increases with
increasing dimensionality. While the energy band gap $E_G$ does
not show a regular trend with dimensionality, the band gap of 2D
monolayer is wider than those of 3D crystals.

\begin{widetext} %
\begin{table*}
\caption{Bonding types; Si-C distances, \textit{d}; lattice constants, \textit{a}; charge transfers, $\delta q$; band gaps, $E_{G}$ with \textit{GW$_{0}$} corrections; and cohesive energies, $E_{c}$ for comparison of SiC polymorphs}

\label{table: comparison}

\centering{}\begin{tabular}{ccccccccccccccccccc}
\hline
structure  &  & Bonding  &  & \textit{d} (\AA{})  &  & \textit{a }(\AA{})  &  & $\delta q$ (e)  &  & $Z_{Si}^{*}$  &  & $Z_{C}^{*}$  &  & $E_{G}$ (eV)  &  & $E_{G}(\mathit{+GW_{0}})$  &  & $E_{C}$ (eV) \tabularnewline
\hline
\hline
Linear Chain  &  & $sp$  &  & 1.649  &  & 3.298  &  & 2.28  &  & 1.72  &  & 6.28  &  & 2.00  &  & - &   & 8.923 \tabularnewline
\hline
Wide Zigzag Chain  &  & $sp$  &  & 1.673  &  & 3.268  &  & 2.15  &  & 1.85  &  & 6.15  &  & 1.18 &  & -  &  & 8.963 \tabularnewline
\hline
2D Honeycomb  &  & $sp^{2}$  &  & 1.786  &  & 3.094  &  & 2.53  &  & 1.47  &  & 6.53  &  & 2.53  &  & 3.90  &  & 11.940 \tabularnewline
\hline
Zincblende  &  & $sp^{3}$  &  & 1.896  &  & 3.096  &  & 2.59  &  & 1.41  &  & 6.59  &  & 1.41  &  & 2.40  &  & 12.939 \tabularnewline
\hline
Wurtzite  &  & $sp^{3}$  &  & 1.893 (3), 1.907 (1)  &  & 3.091  &  & 2.63  &  & 1.37  &  & 6.63  &  & 2.36  &  & 3.32  &  & 12.933 \tabularnewline
\hline
\end{tabular}
\end{table*}
\end{widetext}

\section{Bare and Hydrogen Passivated SiC Nanoribbons}

In this section, we consider bare and hydrogen passivated armchair
SiC nanoribbons. These nanoribbons are specified according to
their widths specified in terms of $N$ number of Si-C basis in
their unit cells. Hence, A-SiCNR($N$) indicates  armchair SiC
nanoribbons having $N$ Si-C pairs in their unit cell. We have
analyzed A-SiCNR (both bare and H-passivated) from $N$=5 to 21.
A-SiCNR's with odd numbered $N$ have reflection symmetry with
respect to their axis. Bare armchair SiC nanoribbons are
ferromagnetic in ideal honeycomb form. However, upon structural
relaxation, reconstruction occurs at the edges resulting in a
considerable gain of energy and the structure becomes non-magnetic.

Band gaps of all A-SiCNR(N) increase by $\sim0.7$ eV upon H-saturation
of dangling bonds. The Si-H and C-H bonds formed after
hydrogenation have lengths of 1.49 \AA{} and 1.09 \AA{},
respectively. The effective charges on Si and C edge atoms changes
after H-passivation. While Si-C bonds at the edges are shorter
than the Si-C bonds in 2D SiC monolayer by 0.09 \AA, upon
H-saturation these bonds get slightly longer, but are still
shorter than the regular Si-C bonds by 0.05 \AA. Hence,
reconstruction of atomic structure at the edge exist in each case,
but is more pronounced in the bare nanoribbons.

\begin{figure}
\centering
\includegraphics[width=8cm]{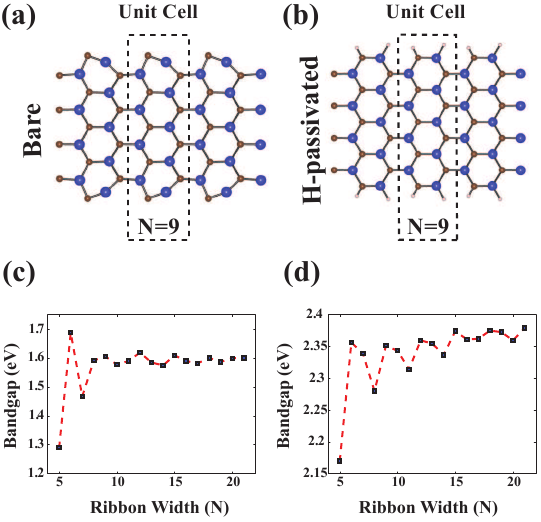}
\caption{(Color online) Energy band structure of the bare (a) and
hydrogen saturated (b) armchair SiC nanoribbons, A-SiCNR($N$) having 9 Si-C pairs ($N=9$) in the unit cell  and isosurfaces of charge densities of selected states at
the $\Gamma$-point of BZ. Zero of energy is set at the Fermi
level. Energy band gap is shaded.}

\label{fig: 9partial}
\end{figure}

Here we consider A-SiCNR(9) as a prototype and examine its band
structure. The bare A-SiCNR($9$) is an indirect band gap
semiconductor. Two bands at the conduction band edge are composed
of edge states, which are split due to edge-edge interactions.
These bands are removed upon H-saturation of dangling bonds of
atoms at the edges. This results in a widening of the band gap. As
for the other edge state band, it is located in the valance band.
This band is also removed upon H-saturation, but the valence band
edge is not affected. The band gap of H-saturated A-SiCNR($9$) is
direct. Energy bands and band decomposed charge densities are
shown in Fig.~\ref{fig: 9partial}.

\begin{figure}
\centering
\includegraphics[width=8.3cm]{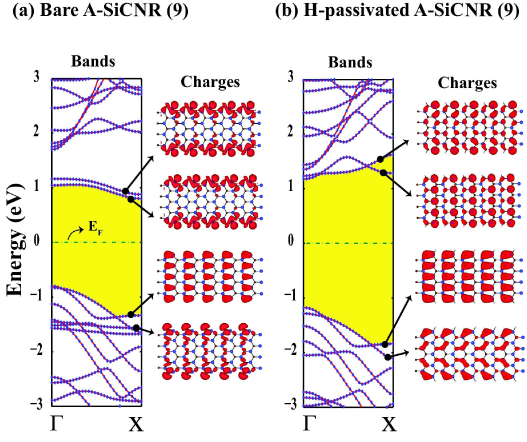}
\caption{(Color online)  Optimized atomic structures for (a) Bare
(b) H-passivated SiC nanoribbons. Large/blue and small/brown balls
represent Si and C atoms respectively. The variations of band gaps of
(c) Bare and (d) H-passivated  SiC nanoribbons with the ribbon width N for $5<N<21$. }

\label{fig: Ribbons}
\end{figure}

The variation of the energy band gap, $E_{G}$, of bare and
H-passivated SiC armchair nanoribbons with the width of the ribbon
$N$ are calculated for 5$\leq$N$\leq$21 as presented in Fig.~\ref{fig: Ribbons}. For bare A-SiCNR's, the band gap is relatively smaller due to edge states as shown in Fig.~\ref{fig: 9partial}; namely $E_{G} \sim$ 1.29 eV for $N$=5, but increases to $\sim$ 2.4 eV for $N$=21. Upon H-passivation of the dangling bonds at the edges, the edge states disappear and the band gap increases and gets direct. For N=5 the direct band gap is around 2.5 eV, but increases with N and eventually becomes 2.38 eV for N=21. Hypothetically, the band gap is
expected to reach the value of 2D planar SiC ($E_G$=2.53 eV
calculated within GGA). The main difference with graphene is that
the band gap of bare armchair graphene nanoribbon decreases with
increasing $N$ and eventually vanish as $N \rightarrow \infty$.
Interestingly, H-saturated armchair SiCNRs with, $N-1$, $N$ and $N+1$, exhibit
a family behavior similar to one revealed in armchair graphene
nanoribbons.\cite{family}  For $N>15$ the variation of $E_{G}$ of H-saturated
A-SiCNR is not significant. However, it should be noted that the
band gap variation of both bare and H-saturated armchair SiC
nanoribbons cannot be reconciled with the quantum confinement effects,
since $E_{G}$ increases with increasing $N$. This is due to other
effects which overcome the quantum confinement effect. The
variation of band gaps of bare and hydrogen passivated
armchair SiC ribbons are presented in Fig.~\ref{fig: Ribbons} (c) and (d). In Ref. \onlinecite{cinsic}, the variation of the band gap with N and resulting family behaviour is in agreement with present results. Whereas the same variation is not presented in  Ref. \onlinecite{cinsic}. The zigzag SiCNRs exhibit interesting magnetic properties and were investigated extensively by earlier works.\cite{cinsic,lou1,lou2} For this reason, the zigzag SiCNRs are examined in the present study.

\section{Vacancy Defects and Antisite}

It has been shown that the vacancy defects have remarkable effects
on 2D graphene honeycomb structure and its
nanoribbons.\cite{esquinazi,Iijima,yazyev,guinea,brey2,topsakal_delik}
Non-magnetic graphene sheets or nanoribbons can attain spin
polarized states due to vacancy defects. We expect that similar
effects of vacancy defects can occur on the electronic and
magnetic properties of SiC honeycomb structure.

\begin{figure}
\centering
\includegraphics[width=8.5cm]{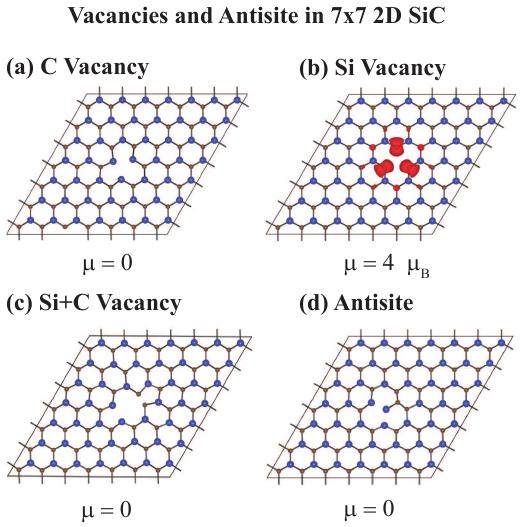}
\caption{(Color online) Optimized atomic structure and magnetic
moment of vacancy defects calculated in a (7$\times$7) supercell
of 2D SiC structure. (a) C vacancy; (b) Si vacancy; (c) C+Si
divacancy; (d) C-Si antisite. In (b) the difference of spin up and
spin down charges are shown. Large/blue and small/brown balls indicate Si and C atoms respectively.}

\label{fig: 7vac}
\end{figure}

\subsection{2D Honeycomb SiC}

The effects of Si and C vacancies, Si+C-divacancy and
C-Si-antisite are treated in  periodically repeating  supercells.
The size of supercell is optimized to allow negligible
defect-defect interaction between adjacent cells. Here the width
of the flat bands derived from the states of periodically
repeating vacancies is taken as the measure of the strength of
vacancy-vacancy coupling. A (7$\times$7) supercell is found to be
suitable, since it leads to rather flat defect bands. At the same
time it is not large and allows us to carry out numerical
calculations within feasible computational time. The flat bands
associated with vacancies can be considered as localized defect
state (if it is in the band gap) or resonance state (if it is in
the band continua). Here we consider localized defect states. Our
results are presented in Fig.~\ref{fig: 7vac} for single C, Si
vacancies, Si+C divacancy, C+Si antisite defects .

A vacancy is generated first by removing a single atom, C or Si
atom from each supercell of the monolayer of SiC as shown in
Fig.~\ref{fig: 7vac} (a) and (b). Subsequently, the atomic
structure is optimized. Single C vacancy in 2D monolayer of SiC is
non-magnetic; Si atoms around vacancy with coordination number 2
are displaced in the transversal direction and do not induce any
magnetic moment.

As for Si vacancy, three C atoms around vacancy remained planar.
Similar to the vacancies in graphene and BN, Si-vacancy induces a
local magnetization in the system. Isovalue surfaces of the
difference between spin up and spin down charge densities i.e.
$\Delta \rho^{\uparrow \downarrow}$ clearly shows a spin
polarization  around the vacancy and a net magnetic moment constructed therefrom. The calculated total magnetic moment is 4 $\mu_{B}$ per supercell. The
Si-vacancy in a repeating (7$\times$7) give rise to defect states
in the band gap.  As for Si+C divacancy in Fig.~\ref{fig: 7vac}
(c), it is again non-magnetic since the spins are paired; two C
atoms around the vacancy choose to make a bond with each other.
The band gap is also modified. Finally, we consider the antisite
defect. The resulting relaxed structure is given in Fig.~\ref{fig:
7vac} (d). Lattice is distorted as C-C bond is shorter than Si-Si
bond in the antisite case. It is noted that the calculated
magnetic moments for single Si- and C-vacancy do not agree with
Lieb's theorem,\cite{lieb} which normally predicts $1 \mu_B$ net
magnetic moment both for Si and C vacancies in Fig.~\ref{fig:
7vac}. We attribute the discrepancy between the results of first
principles calculations and Lieb's theorem to the structural
relaxation occurred after the generation of vacancy and
significant charge transfer from Si to C. The localized electronic
states associated with the vacancy defects and antisite are deduced from the band structure calculations as
presented in Table~\ref{table: 7vac}, where the energies are
given from the top of the valance band.

\begin{table}
\caption{Magnetic moments and positions of defect-induced state energies relative to the top of the valance band. Spin-up($\uparrow$) and spin-down($\downarrow$) states are indicated. E (F) indicate whether the defect state is empty (full-occupied).}

\label{table: 7vac}

\centering{}\begin{tabular}{ccccc}
\hline
 & Si+C Vacancy  & C-Vacancy  & Si-Vacancy  & Antisite\tabularnewline
\hline
\hline
$\mu$ & 0  & 0  & 4 $\mu_{B}$  & 0 \tabularnewline
\hline
$E_{1}$  & 0.49 (F)  & 0.15 (F)  & 0.09 (F $\uparrow$)  & 0.59 (F) \tabularnewline
\hline
$E_{2}$  & 1.40 (E)  & 1.82 (E)  & 0.11 (F $\uparrow$)  & 2.12 (E) \tabularnewline
\hline
$E_{3}$  & 2.45 (E)  & 1.95 (E)  & 0.31 (F $\downarrow$)  & - \tabularnewline
\hline
$E_{4}$  & -  & -  & 0.33 (F $\downarrow$)  & - \tabularnewline
\hline
$E_{5}$  & -  & -  & 0.38 (E $\downarrow$)  & - \tabularnewline
\hline
$E_{6}$  & -  & -  & 1.47 (E $\downarrow$)  & - \tabularnewline
\hline
$E_{7}$  & -  & -  & 1.49 (E $\downarrow$)  & - \tabularnewline
\hline
\end{tabular}
\end{table}

\subsection{Vacancy defects in SiC Nanoribbons}

\begin{figure}
\centering
\includegraphics[width=9.2cm]{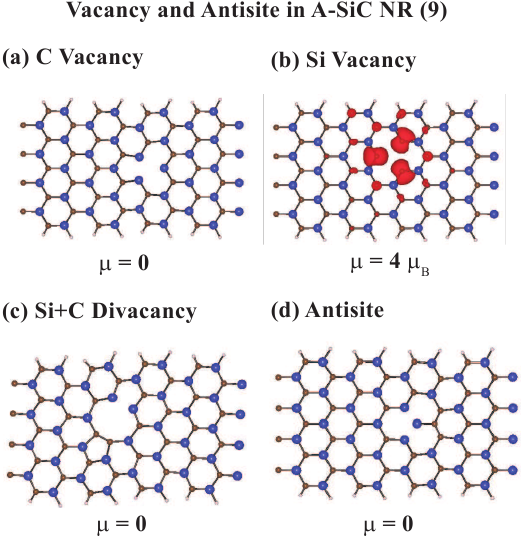}
\caption{(Color online) Magnetic moment of vacancy defects
calculated in a (4$\times$1) supercell of quasi 1D SiC armchair
nanoribbon with $N$=9, i.e. A-SiCNR(9). Dangling bonds at both
edges are saturated by hydrogen atoms. (a) C vacancy; (b) Si
vacancy; (c) Si+C divacancy; (d) C-Si antisite. In (b) the
difference of spin up and spin down charges are shown while others are non-magnetic. Large/blue and small/brown balls represent Si and C atoms respectively.}

\label{fig: ribbvac}
\end {figure}

The effects of the vacancy defects on the electronic and magnetic
properties are treated for of H-passivated A-SiCNR(9) using a
(4$\times$1) repeating supercell. Our main motivation was to
investigate what differences would occur in a ribbon. Optimized
structures, calculated total magnetic moments are presented in
Fig.~\ref{fig: ribbvac}. Overall effects of vacancy, divacancy and
antisite effects are similar to those in 2D SiC monolayer
structure except in the antisite case  the exchanged Si atom moves
out of the plane about 0.9 \AA{}. One localized state below the
conduction band edge and one other state above the valence band
edge occur due to antisite as donor and acceptor states,
respectively.

\section{Functionalization of SiC honeycomb structure by adatoms}

Specific adatoms can bind to 2D SiC monolayer with significant
binding energies. Adatom adsorption or decoration, as well as
substitution for Si or C atoms in the honeycomb structure by
foreign atoms can modify the properties of 2D SiC monolayer and
its nanoribbons. This way,  SiC honeycomb structures can be
functionalized. Adatom adsorption and substitution are considered
within the periodically repeating (7$\times$7) supercell geometry
to minimize the interaction between them. The calculations performed in a larger supercell (10$\times$10) have also showed similar results which is an indication that the coupling between adjacent defects is negligible.

\subsection{Adatom adsorption}

\begin{table}
\caption{$E_{B}$, total magnetic moments $\mu$, optimized heights of adatoms
from the 2D monolayer of SiC, \textit{h} and localized states of occurring
in the band gap are given for each type of adatom. Empty (E), full
(F), spin-up and spin-down states are indicated. Energies of adatom
induced localized states are given relative to the top of the valence
band. }

\label{table:adsorp}

\centering{}\begin{tabular}{c|c|c|c|c}
\hline
 & Co  & Fe  & P  & Ti \tabularnewline
\hline
\hline
Position  & HS  & HS  & BS  & TS \tabularnewline
\hline
$E_{B}$ (eV)  & 2.099  & 1.928  & 1.758  & 2.605 \tabularnewline
\hline
$\mu$ ($\mu_{B}$)  & 1  & 2  & 1  & 2 \tabularnewline
\hline
\textit{h} (\AA{})  & 1.392  & 1.465  & 1.743  & 1.355 \tabularnewline
\hline
\hline
$E_{1}$  & 0.13 (F $\uparrow$)  & 0.30 (F $\uparrow$)  & 0.17 (F $\uparrow$)  & 0.43 (F $\uparrow$) \tabularnewline
\hline
$E_{2}$  & 0.23 (F $\uparrow$)  & 0.31 (F $\uparrow$)  & 0.16 (F $\uparrow$)  &  0.60 (F $\downarrow$) \tabularnewline
\hline
$E_{3}$  & 0.33 (F $\downarrow$)  & 0.82 (F $\downarrow$)  & 0.20 (F $\downarrow$)  & 1.62 (F $\uparrow$) \tabularnewline
\hline
$E_{4}$  & 0.41 (F $\downarrow$)  & 0.82 (F $\downarrow$)  & 0.20 (F $\uparrow$)  & 1.62 (F $\uparrow$) \tabularnewline
\hline
$E_{5}$  & 0.67 (F $\uparrow$)  & 1.17 (F $\downarrow$)  & 0.31 (F $\downarrow$)  & 1.99 (E $\uparrow$) \tabularnewline
\hline
$E_{6}$  & 0.80 (F $\downarrow$)  & 1.93 (E $\uparrow$)  & 0.33 (E $\downarrow$)  & 2.28 (E $\uparrow$) \tabularnewline
\hline
$E_{7}$  & 1.26 (E $\downarrow$)  & 2.20 (E $\downarrow$)  & -  & 2.40 (E $\downarrow$) \tabularnewline
\hline
\end{tabular}
\end{table}

As for  adatom adsorption, we considered Al, Co, Fe, N, P, Ti by placing each of
them on four different positions in the (7$\times$7) monolayer of
SiC and then by fully relaxing the whole system. The initial
positions of the adsorption are on top of silicon atom (TS), on
top of carbon atom (TC), at the center of hexagon (HS), above the
middle of the Si-C bond (BS). The distance between adatoms is ~ 12
\AA. Spin polarized calculations are carried out to determine the
binding structure and binding energy. Whether the adatoms are
bound to the surface are examined by calculating the binding
energies of these six different individual atoms in terms of the
calculated total energies as $E_B={E_{T}[SiC+adatom]} -
E_{T}[SiC(bare)] - E_{T}[adatom]$. We found that all of these
adatoms are bound with a significant energy, which is larger than
1 eV.  Flat bands indicate that states induced by the adatoms are
rather localized and hence adatom-adatom interactions are
negligible. Therefore, adatoms treated here in supercell geometry
can represent single (isolated) adatom. Optimized adsorption
sites, total magnetic moments, heights from the SiC plane are
given in Table~\ref{table:adsorp}.

\subsection{Substitution of Si and C by foreign atoms}

Here we examined the substitution of single Si or C atoms in 2D
SiC honeycomb structure by various foreign atoms. Namely B and N
substituting C atom; Al, As, Ga, P substituting Si atom. Similar
to adatom calculations, the substitution process is treated within
periodically repeating (7$\times$7) supercell. Because of periodic
boundary condition the localized states appear as flat bands. We
found that B, N, As and P atoms have a net magnetic moment $\mu$=
1 $\mu_{B}$. Ga and Al do not create any spin polarization. In
Table~\ref{table:subs}, the substitutional impurity states in the band gap of 2D SiC and resulting net magnetic moments are given. The
substitutional foreign atoms, namely B, N, As, and P have either 3
or 5 valence electrons and hence destroy the spin pairing in the
perfect honeycomb structure. The substitution of C and Si by these
atoms gives rise to a net magnetic moment of 1 $\mu_B$. As a
result, 2D SiC can be magnetized without doping by transition
metal elements.

\begin{table}
\caption{Energies of defects states occurring in the band gap are
given for each type of  substituted atom. Empty (E), full (F),
spin-up and spin-down states are indicated. Energies are measured
from the top of valence band.} \label{table:subs}

\begin{center}
\begin{tabular}{ccccc}
\hline  \hline

       & B & N & As & P \\ \hline \hline
$\mu(\mu_{B})$ & 1  & 1  & 1  & 1  \\ \hline
$E_{1}$ & 0.23 (F, $\uparrow$) &   1.47 (F, $\uparrow$) & 0.75 (F, $\uparrow$) &  1.18 (F, $\uparrow$)            \\ \hline
$E_{2}$ & 0.69 (E, $\downarrow$) &     2.37 (E, $\downarrow$) &   1.18 (E, $\downarrow$) &    1.77 (E, $\downarrow$)          \\ \hline

\end{tabular}
\end{center}
\end{table}

\section{Discussion and Conclusions}

We present a study on 1D SiC chains, 2D monolayer of SiC in
honeycomb structure and its armchair nanoribbons, 3D bulk SiC. We
carried out stability analysis of those materials. Two dimensional
monolayer of SiC is an ionic compound with charge transfer from
silicon atoms to carbon atoms and has a hexagonal lattice forming
a planar honeycomb structure. The calculation of phonon modes
results in all positive frequencies in BZ and indicates the
stability of the planar honeycomb structure. This situation is in
compliance with the previous works,\cite{silicene,hasan} where the
honeycomb structures of binary compounds of group IV elements or
III-V compounds are stable in planar geometry if they have an
element from the first row, such as B, C, N. The 2D SiC is a non-magnetic
wide band gap semiconductor. However, it acquires net magnetic
moment when a single Si-vacancy defect is created or Si and C atoms of the honeycomb structure are substituted by As, P, B and N.  Single C-vacancy, Si+C divacancy and Si-C antisite defects does not give rise to any magnetic moment in the system.
It is shown that 2D SiC can be functionalized through Si-vacancy
and adatom adsorption or substitution of Si C by foreign atoms.

Armchair SiC nanoribbons are found to be nonmagnetic
semiconductors. We revealed the variation of band gap with the
width of both bare and H-passivated nanoribbons. The variation of
the band gap exhibits also family behavior. However,
the band gap of armchair nanoribbons are smaller than the band gap of 2D SiC  for small N. Therefore the confinement effect seen in narrow graphene armchair nanoribbons does not occur here.

In conclusion, our state-of-the-art first-principles calculations
demonstrate that 2D SiC monolayer is stable in honeycomb
structure, and hence it has a strong chance to be  synthesized in future. Above
results indicate that bare and H-passivated SiC sheets and
armchair nanoribbons can present interesting properties which can
be utilized in nanotechnology. Creating defects through vacancies,
doping through adatoms and substitutional impurities can
functionalize SiC honeycomb structure and hence add new magnetic
and electronic properties.

\begin{acknowledgments}
We would like to thank Professor Dario Alf\`{e} for his valuable
discussions. Part of the computations have been carried out with
the service provided by UYBHM at Istanbul Technical University
through a Grant No. 2-024-2007. This work is partially supported
by TUBA, Academy of Science of Turkey.
\end{acknowledgments}

\end{document}